\documentclass[%
 reprint,
superscriptaddress,
 amsmath,amssymb,
 aps,
]{revtex4-2}

\usepackage{graphicx}
\graphicspath{ {Figures/} }
\usepackage{dcolumn}
\usepackage{bm}
\usepackage[hyperindex,breaklinks]{hyperref}
\hypersetup{
    colorlinks=true,
    linkcolor=red,
    citecolor=blue,
    urlcolor=blue
}
\usepackage{breakurl}
\usepackage[mathlines]{lineno}
\usepackage[caption=false]{subfig}
\usepackage{booktabs}
\usepackage{ulem} 

\newcommand{\nuebar}{$\overline{\nu}_{e}$ }
\newcommand{\nuebars}{$\overline{\nu}_{e}$'s }
\newcommand{\alphan}{($\alpha, n$) }
\newcommand{\alphap}{$\alpha$-$p$ }

\begin{document}

\preprint{APS/123-QED}

\title{Measurement of reactor antineutrino oscillations with 1.46 ktonne-years of data at SNO+}

\author{ M.\,Abreu }
\affiliation{ Laborat\'{o}rio de Instrumenta\c{c}\~{a}o e  F\'{\i}sica Experimental de Part\'{\i}culas (LIP), Av. Prof. Gama Pinto, 2, 1649-003, Lisboa, Portugal }
\affiliation{ Universidade de Lisboa, Instituto Superior T\'{e}cnico (IST), Departamento de F\'{\i}sica, Av. Rovisco Pais, 1049-001 Lisboa, Portugal }
\author{ A.\,Allega }
\affiliation{ Queen's University, Department of Physics, Engineering Physics \& Astronomy, Kingston, ON K7L 3N6, Canada }
\author{ M.\,R.\,Anderson }
\affiliation{ Queen's University, Department of Physics, Engineering Physics \& Astronomy, Kingston, ON K7L 3N6, Canada }
\author{ S.\,Andringa }
\affiliation{ Laborat\'{o}rio de Instrumenta\c{c}\~{a}o e  F\'{\i}sica Experimental de Part\'{\i}culas (LIP), Av. Prof. Gama Pinto, 2, 1649-003, Lisboa, Portugal }
\author{ D.\,M.\,Asner }
\affiliation{ SNOLAB, Creighton Mine \#9, 1039 Regional Road 24, Sudbury, ON P3Y 1N2, Canada }
\affiliation{ Brookhaven National Laboratory, P.O. Box 5000, Upton, NY 11973-500, USA }
\author{ D.\,J.\,Auty }
\affiliation{ University of Alberta, Department of Physics, 4-181 CCIS,  Edmonton, AB T6G 2E1, Canada }
\author{ A.\,Bacon }
\affiliation{ University of Pennsylvania, Department of Physics \& Astronomy, 209 South 33rd Street, Philadelphia, PA 19104-6396, USA }
\author{ T.\,Baltazar }
\affiliation{ Laborat\'{o}rio de Instrumenta\c{c}\~{a}o e  F\'{\i}sica Experimental de Part\'{\i}culas (LIP), Av. Prof. Gama Pinto, 2, 1649-003, Lisboa, Portugal }
\affiliation{ Universidade de Lisboa, Instituto Superior T\'{e}cnico (IST), Departamento de F\'{\i}sica, Av. Rovisco Pais, 1049-001 Lisboa, Portugal }
\author{ F.\,Bar\~{a}o }
\affiliation{ Laborat\'{o}rio de Instrumenta\c{c}\~{a}o e  F\'{\i}sica Experimental de Part\'{\i}culas (LIP), Av. Prof. Gama Pinto, 2, 1649-003, Lisboa, Portugal }
\affiliation{ Universidade de Lisboa, Instituto Superior T\'{e}cnico (IST), Departamento de F\'{\i}sica, Av. Rovisco Pais, 1049-001 Lisboa, Portugal }
\author{ N.\,Barros }
\affiliation{ LIP – Laboratório de Instrumentação e Física Experimental de Partículas, Escola de Ciências, Campus de Gualtar, Universidade do Minho, 4701-057 Braga, Portugal }
\affiliation{ Departamento de Fisica, Universidade do Minho, 4710-057 Braga, Portugal }
% ...
\collaboration{ The SNO+ Collaboration }
\noaffiliation


\begin{abstract}
\newpage
The SNO+ Collaboration reports new results on reactor antineutrino oscillations using data acquired from May 2022 through July 2025. 
The spectral analysis of a flux dominated by nuclear reactors at 240, 340, and 350 kilometers yields the mass-squared difference $\Delta m^2_{21}=(7.91^{+0.22}_{-0.25})\times 10^{-5}$~eV$^2$.  
This result is compatible with and approaches the precision of the only other long-baseline reactor antineutrino measurement, by KamLAND. 
Combining these measurements, along with those from solar neutrino experiments, the global values of the neutrino mixing parameters become: $\Delta m^2_{21}$ = $(7.59\pm0.17)\times 10^{-5}$ eV$^2$ and $\sin^2{\theta_{12}}=0.310\pm0.012$. 
The analysis of geoneutrinos at SNO+ is also improved, with a measured signal of 48$^{+14}_{-12}$ TNU.
\end{abstract}

\maketitle


\textit{Introduction}~~\textemdash~~
Neutrino oscillations can be described by three mixing angles, $\theta_{12}$, $\theta_{23}$, $\theta_{13}$, and two mass-squared differences, $\Delta m_{21}^2$ and $\Delta m_{32}^2$. These parameters have been measured to have consistent values using rates and spectral features from various neutrino and antineutrino sources. The smaller difference, $\Delta m_{21}^2$, has been measured using electron antineutrinos from commercial nuclear reactors at long and medium baselines~\cite{kamland_on_off, sno+ppo, JUNO:2025gmd} and electron neutrinos from the Sun~\cite{Super-Kamiokande_solar}.  

SNO+ is the second experiment to use long-baseline reactor antineutrinos to measure neutrino oscillations, doing so at distinct baselines, which lead to distinct spectral features.  
The results reported in this Letter are obtained from data collected from May 2022 through July 2025, including those used in a previous analysis~\cite{sno+ppo}. 
The current dataset is five times larger and improves the precision of the measured oscillation parameters and geoneutrino flux by a factor of two. 
In the following, we introduce a new parameter to discriminate antineutrino signals from the dominant \alphan background. This is an evolution of a preliminary study~\cite{snoplus_partial_antinu} and is applied here for the first time, further improving the geoneutrino measurement. 

\smallskip
\textit{Data}~~\textemdash~~
The SNO+ detector is located 2.1~km underground at SNOLAB in Ontario, Canada.  It is equipped with 9362 photomultiplier tubes (PMTs), which view a 6-m radius acrylic vessel (AV) filled with about 780 tonnes of scintillator. Ultrapure water shields the AV from radioactivity in the surrounding rock and in the PMTs, which are 8.5~m from the center. 
More detailed descriptions of the detector and the scintillator can be found in Refs.~\cite{sno+detector,sno+scintillator}.

In a first phase of the experiment, the AV was filled with ultrapure water, which was later replaced by an LAB-based scintillator filled in from the top of the AV as the water was removed from the bottom. Filling was paused when the scintillator reached 75~cm above the equator of the AV, providing the stable dataset reported in Ref.~\cite{snoplus_partial_antinu}. 
The first analysis of data collected with a fully-filled AV, and with the primary fluor PPO at a concentration of 2.2~g/L, was published in Ref.~\cite{sno+ppo}. 
That dataset used 134.4 days, or 0.29 ktonne-years, of data and is now extended, as the criteria for the fraction of online PMTs have been relaxed due to an improved accounting of offline channels in the energy reconstruction.  Here, it is denoted as dataset I. 
Dataset II corresponds to the period after the addition of 2.2~mg/L of the wavelength shifter bis-MSB, which increased the number of collected photons by about 50\%.  
The relevant characteristics of these datasets are summarized in Table~\ref{tab:datasets}. 

\begin{table}[tbp]
    \centering
    \caption{\label{tab:datasets} Characteristics of the two datasets, including photon collection and the resulting energy resolution for 1-MeV electrons at the center of the scintillator volume.  The bottom three rows are evaluations of energy-related systematic uncertainties.  See text for details.}
\begin{tabular}{lcc}
\toprule \toprule
Dataset & I & II \\
\toprule
Scintillator & LAB+PPO & + bis-MSB\\
Dates & 2022-2023 & 2023-2025\\
Livetime [days] &  245.8 & 439.4  \\
Photon collection [PMTs/MeV] & 210  & 320 \\
Energy resolution & 6.5\% & 5.0\% \\
Energy resolution unc. & 4.4\% & 4.9\% \\
Energy scale unc. - linear & 1.1\% & 1.0\% \\
Energy scale unc. - $k_B$ & 5.4\% & 5.4\% \\
\bottomrule \bottomrule
\end{tabular}
\end{table}

Data were first selected with a set of criteria designed to reduce instrumental backgrounds as well as backgrounds induced by the passage of a muon.  
Data were vetoed for 20~s after an identified muon and for 1~s after events with at least 3000 PMTs hit, corresponding to unidentified or glancing muons, or high-energy atmospheric neutrinos.  Data were also vetoed for 10~$\mu$s after events with 3 or more outward-looking PMTs hit, to mitigate potential signals from externally-produced fast neutrons.  Given the low muon rate at SNO+ ($\approx$3 per hour), these vetoes result in a 2\% reduction in livetime and a negligible muon-induced background.  

The observable energy of an event is, to first order, proportional to the number of PMTs that detect photons.  
The differences in photon collection between the datasets in Table~\ref{tab:datasets} are due not only to changes in scintillator photon yield, but also to differences in optical absorption and a general increase in the number of offline PMTs with time.  
The energy reconstruction accounts for the optical properties, as well as the reconstructed position of an event and the known position of each offline PMT.  The probability of multiple incident photons on the PMTs is accounted for analytically. 

Electron antineutrinos with energies above 1.8~MeV are detected through the inverse beta decay (IBD) process: \nuebar$ +~p \to e^+ + n$. The antineutrino energy $E_{\text{\nuebar}}$ is transferred primarily to the positron, which promptly deposits its energy and annihilates with an electron in the medium, resulting in an observable energy of $\approx(E_{\text{\nuebar}}-~0.8)$~MeV.  The neutron thermalizes and is captured by a proton, producing a 2.2-MeV $\gamma$, after a characteristic $\approx$210 $\mu s$ random walk with a maximum distance of around 1~m. 

\begin{figure}[tbp]
\caption{\label{fig:position_plots} Positions of prompt and delayed events selected in the full dataset, within the 5.7~m fiducial volume, projected onto ($z$, $x^2 + y^2$) to preserve equal volume elements. Dotted lines connect prompt-delayed pairs.}
\includegraphics[width=1.0\columnwidth]{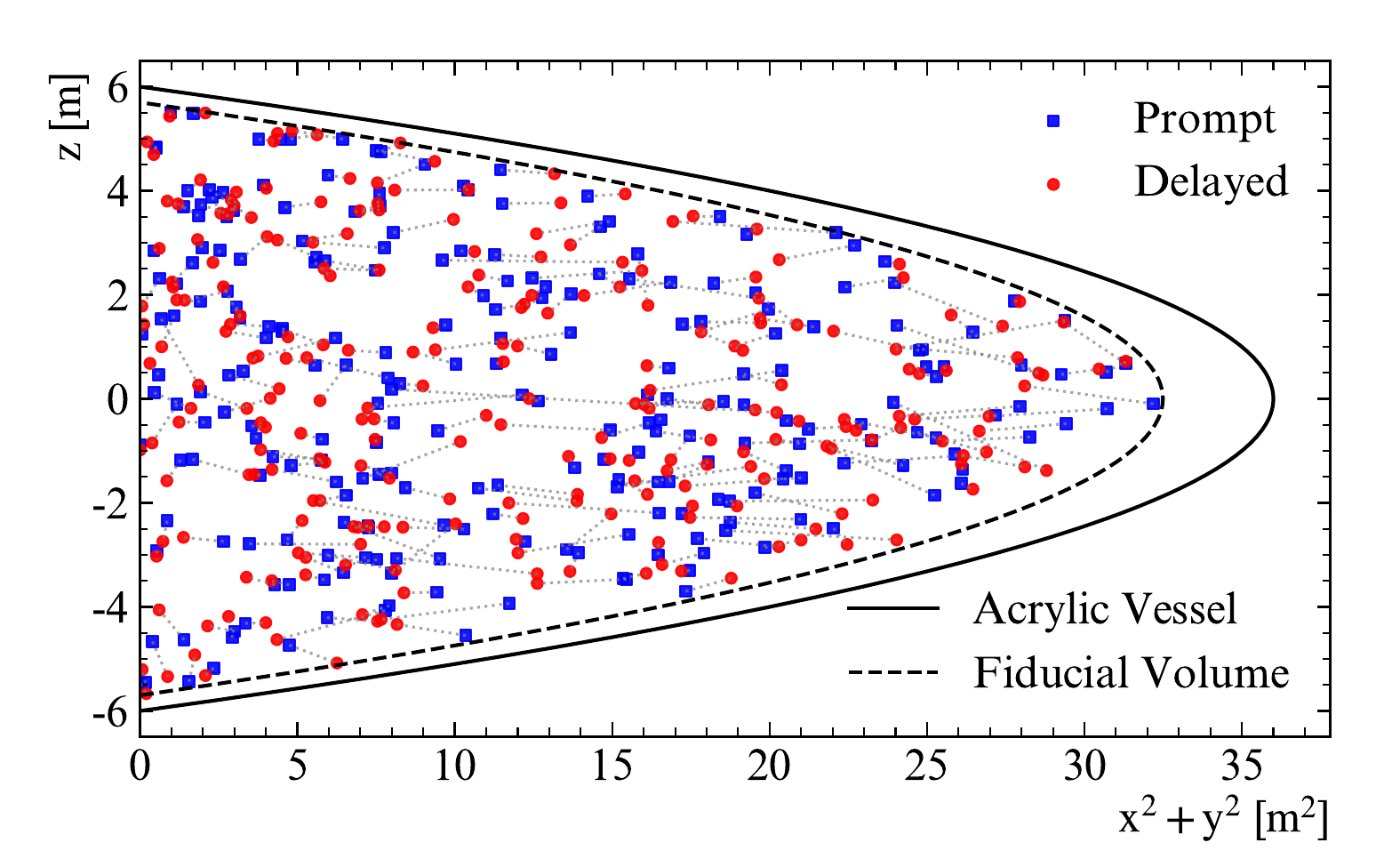}
\end{figure}

IBDs were selected within a fiducial volume of 5.7~m, which is chosen to reduce backgrounds from the AV. Prompt events with an energy between 0.9~MeV and 8~MeV were kept if they were followed by a delayed event with an energy between 1.85 MeV and 2.5~MeV, within a time $\Delta t <$ 2~ms, and within a distance $\Delta r$ $<$ 2.5~m.  Coincidences with more than one candidate for a prompt or delayed event were vetoed.  A likelihood ratio that combines the time and distance between prompt and delayed events ($\Delta t$ and $\Delta r$), and the delayed event energy, was used to compare IBDs and random coincidences arising from radioactivity.  The latter were predicted from data by randomly pairing single events observed during $\approx$1-hour periods. 
A cut on this likelihood ratio reduced random coincidences to negligible levels of $\mathcal{O}(10^{-1})$ counts in each dataset. 
The spatial distribution of all coincidences selected with these criteria is uniform throughout the fiducial volume, as seen in Fig.~\ref{fig:position_plots}.  
Furthermore, since the IBD signals and dominant \alphan background all have neutron-capture delayed events, the $\Delta t$ distribution follows an exponential distribution.  We found a fitted time constant of $\tau=(217\pm15)~\mu$s, in agreement with the expected 210~$\mu$s.  

\smallskip
\textit{Calibrations}~~\textemdash~~
Intrinsic $^{214}$Bi$^{214}$Po decays were used to calibrate the scintillation time profile, photon yield, and quenching, for $\beta$'s and $\alpha$'s, as described in the previous analyses~\cite{snoplus_partial_antinu,sno+ppo}. The residual nonuniformity of the reconstructed energy scale was also 
corrected and quantified in a very similar way, resulting in the uncertainties in Table~\ref{tab:datasets}.
The nonlinear scaling due to Birks’ law was used to apply an uncertainty in the Birks constant $k_B$ as done in the previous analyses. 

A first deployment of a calibration source ($^{241}$Am-$^9$Be) in the scintillator took place in July 2025.  
$^9$Be$(\alpha,n)^{12}$C interactions in the AmBe source produce neutrons with energies up to 10~MeV, which scatter protons in the scintillator. In this way, the AmBe source mimics the proton scatters from high-energy neutrons produced by $^{13}$C$(\alpha,n)^{16}$O interactions in the scintillator, and so it is used to evaluate the use of a new event classifier for the dominant \alphan background. 

The \alphan classifier distinguishes the multiple-proton scattering signals of MeV-scale neutrons that are characteristic of \alphan prompt signals below 3.5~MeV, from the prompt positrons created in IBDs. 
It is built as a Fischer discriminant, using the distribution of time-of-flight corrected PMT hit times as well as the reconstructed radial event position. 
In the pure samples of $^{214}$Bi$^{214}$Po coincidences used for the scintillator calibrations, the classifier correctly assigned the $\beta$'s to be more positron-like than the $\alpha$'s. 
The classification values in simulation and data diverged at high radial positions; 
therefore, an empirical correction was developed to shift the classification in simulation to match the data as a function of 
the distance to the center of the detector. 
The residual discrepancy for $^{214}$Bi was used to assign a 2.9\% uncertainty to the IBD selection efficiency. 
The corrected distributions of \alphan classifier values for the $^{214}$Bi$^{214}$Po $\alpha$'s and $\beta$'s, which illustrate the agreement between the data and simulations, are shown in Supplemental Material~\cite{supp}. 

Regarding AmBe prompt events, the \alphan classifier was found to depend on energy differently between the data and simulation, leading to an energy-dependent systematic uncertainty on the efficiency with which \alphan events were rejected; namely, an upper limit for the differences between the observed multiple-proton scattering signals and those in the simulations.  Since dataset I does not have an associated AmBe deployment, the uncertainty was assumed to have the same energy dependence, to be of the same relative magnitude, and to be uncorrelated with that for dataset II.

The selection threshold for the classification value of prompt events was determined separately for each dataset, optimizing the signal to background ratio with IBD and \alphan simulations. 
The selection thresholds are expected to preserve about 90\% of geonenutrino IBDs, and a similar fraction of the 60\% of reactor IBDs with prompt energies below 3.5~MeV. 
Only 17\% and 8\% of \alphan events below 3.5~MeV are expected to be retained in datasets I and II, respectively. The classifier is more powerful in dataset II due to the higher photon collection.

\smallskip
\textit{Reactor Antineutrinos}~~\textemdash~~
Nuclear reactors produce electron antineutrinos with energies up to $\approx$10 MeV. The flux is dominated by four fissile isotopes ($^{235}$U, $^{238}$U, $^{239}$Pu, $^{241}$Pu), each with a similar but distinct antineutrino spectrum and effective energy release per fission. At SNOLAB, $>$99\% of the flux originates from the $\approx$100 active reactor cores in North America, with $\approx$60\% from three complexes in Ontario, where CANDU pressurized heavy water reactors (PHWRs) are used. 
The average fractions of the four isotopes in these reactors are a little different from the more common pressurized/boiling water reactors (PWRs/BWRs): (0.52, 0.05, 0.42, 0.01) vs. (0.568, 0.078, 0.297, 0.057)~\cite{sno+ppo}. 
The thermal power of each core was tracked daily using publicly available data for the electrical output, from IESO in Ontario~\cite{ieso} and NRC in the USA~\cite{nrc}.  These data were cross-checked with the monthly averages from IAEA~\cite{iaea}, leading to a small uncertainty of around 0.2\%.  

The main fissile isotopes, $^{235}$U and $^{239}$Pu, produce roughly 90\% of the \nuebars and their spectra have been obtained by a joint fit to data from Daya Bay and PROSPECT~\cite{daya_bay_prospect_spectrum}. Predictive models are used for $^{241}$Pu~\cite{huber2011determination} and $^{238}$U~\cite{mueller2011improved}. 
These spectra were used to determine the expected fluxes for both types of reactors.  
The covariance matrices provided in Ref.~\cite{daya_bay_prospect_spectrum} were integrated and found to dominate the total flux uncertainty, at about 3\%.  
Following the prescription of Ref.~\cite{an2021antineutrino}, bin-by-bin correlations in the reactor IBD spectra were calculated. Due to their high, positive values, their effect on the fitted uncertainty of $\Delta m^2_{21}$ was found to be negligible, and so are not included in this analysis.

Following the estimations in Ref.~\cite{DayaBay2019}, we also propagated the subdominant uncertainties for the energy released per fission, time evolution of the isotopes, and correlations between the spectra, resulting in a combined uncertainty of 4.0\% for PWRs and 4.3\% for PHWRs. 
Finally, taking into account the subdominant uncertainties in proton number and position reconstruction as in the previous analysis~\cite{sno+ppo}, a common systematic of 4.3\% was conservatively assigned to all reactor predictions.

\begin{figure}[tbp]
    \centering
    \caption{Energy distribution of prompt events and best-fit predictions.  Data are given wider bins to aid visual comparison.  Upper panel: fit with $\Delta m^2_{21}$ and $\sin^2\theta_{12}$ unconstrained.  Lower panel: with constraints and the cut on the \alphan classifier, which is applied only below 3.5~MeV (dashed line).  The reactor IBD expectation is comprised of three distinct groups: Bruce at 240~km, Pickering + Darlington (P + D) at 340 and 350~km, and all other reactors (World).}
    \includegraphics[width=1.0\columnwidth]{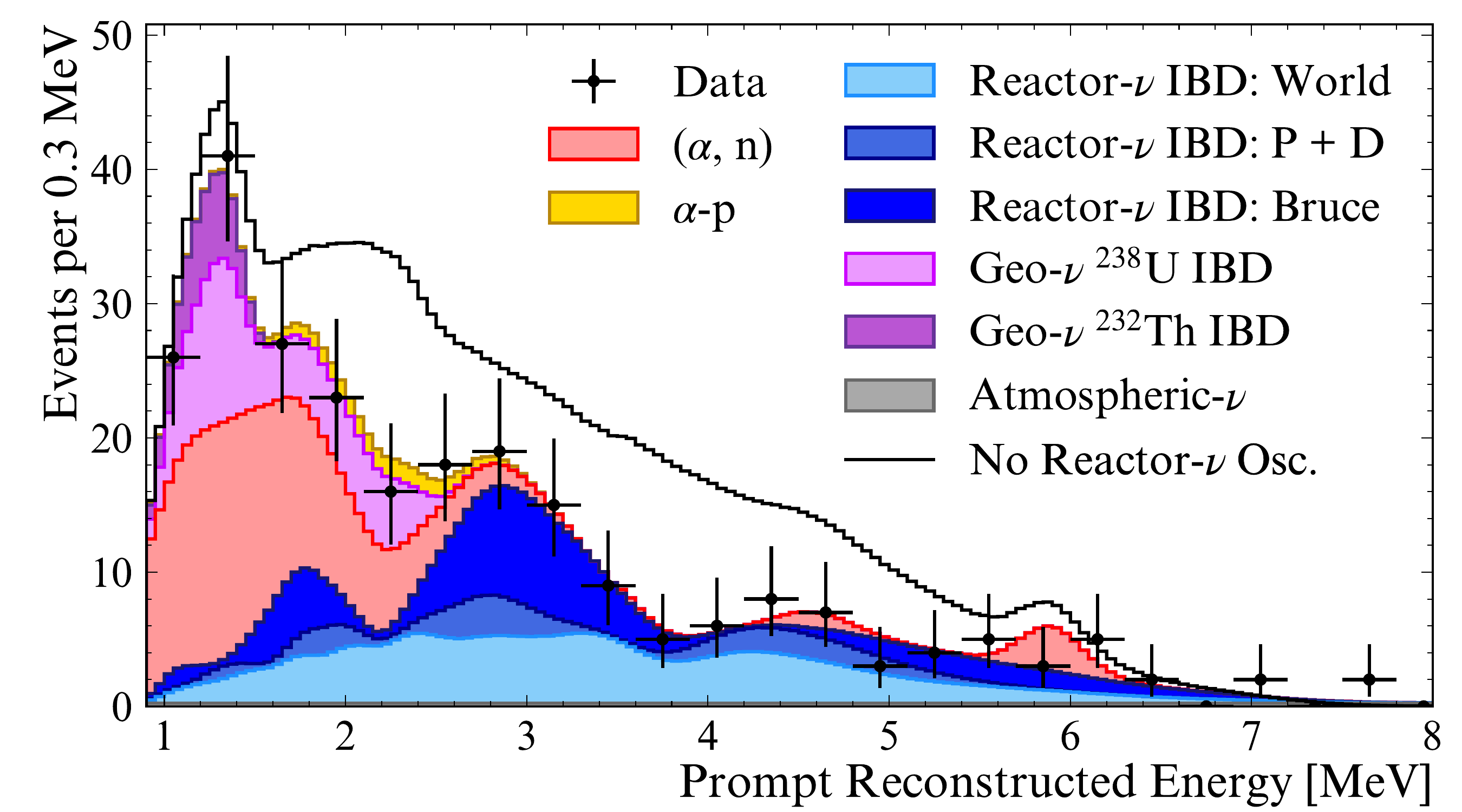}
    \includegraphics[width=1.0\columnwidth]{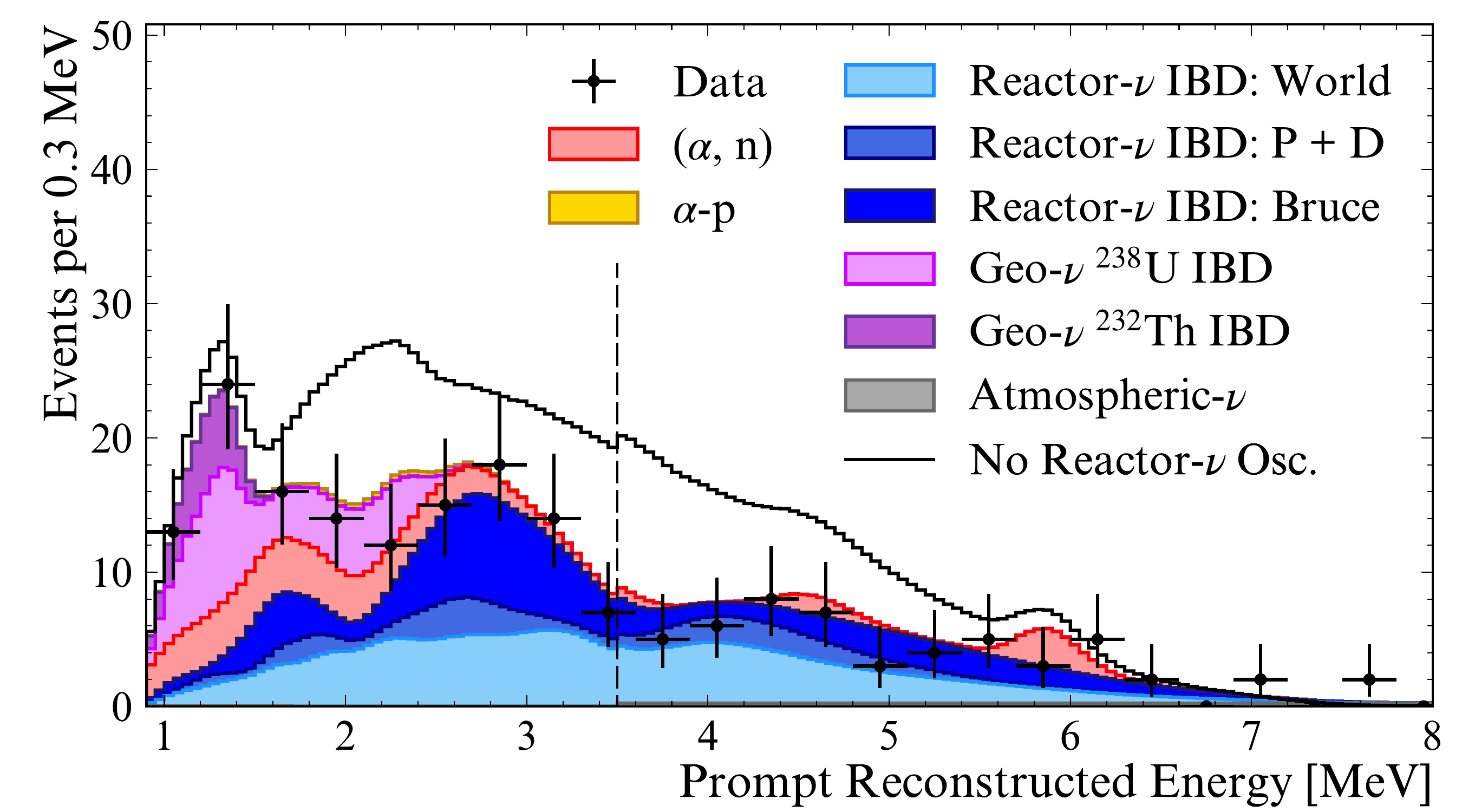}
    \label{fig:Eprompt}
\end{figure}

\smallskip
\textit{Neutrino Oscillation}~~\textemdash~~
The survival probability of an electron (anti)neutrino in vacuum is 
\begin{equation*}
\begin{aligned}
P_{ee} = 1
& \left. -\cos^{4}\theta_{13}\sin^{2} 2\theta_{12}\sin^{2}\Delta_{21}\right.\\
& \left. -\sin^{2}2\theta_{13}(\cos^{2}\theta_{12}\sin^{2}\Delta_{31} + \sin^{2}\theta_{12}\sin^{2}\Delta_{32}), \right.\\
\end{aligned}
\end{equation*}
where $\Delta_{ij} \equiv 1.267\Delta m_{ij}^{2}L/E$, $L$ [m] is the distance traveled by the neutrino, $E$ [MeV] is the energy of the neutrino, and $\Delta m_{ij}^{2} \equiv m_i^2-m_j^2$ [eV$^2]$ is the difference between the squares of the masses of neutrino mass eigenstates $i$ and $j$. 
The oscillation pattern in the energy spectrum of IBD prompt events arises from $\Delta m^2_{21}$, as illustrated in Fig. 3 of Ref.~\cite{snoplus_partial_antinu}.  The nearest reactor complex, Bruce at 240~km from SNOLAB, provides $\approx$44\% of reactor IBDs, resulting in a relatively large amplitude in the observed oscillation pattern.  There are also supporting contributions from the two other complexes in Ontario at 340~km and 350~km, which are shown in Fig.~\ref{fig:Eprompt}.

Considering that more than 95\% of incident \nuebars travel through the North American crust, the effects of matter are included according to Ref.~\cite{mattOsc}, assuming a constant electron density of 8.13~$\times$10$^{23}$~cm$^{-3}$. 
Relative to vacuum oscillations, this increases the number of IBDs by less than 1\% and induces a $\mathcal{O}(1\%)$ increase in the effective $\Delta m_{21}^2$. 
Using the measured oscillation parameters in Ref.~\cite{PDG2025}, the expected rate of reactor IBDs is around 100 per year in the full AV volume. 

\smallskip
\textit{Geoneutrinos}~~\textemdash~~
The natural $^{238}$U and $^{232}$Th in the Earth's crust and mantle each undergo a series of beta decays that produce \nuebars with energies up to 3.3 MeV.  Dedicated geological measurements have been made to obtain a model for the geoneutrino fluxes at SNOLAB, where the contribution from the thicker North American crust is higher than at the locations of previous measurements~\cite{geo-local}. A detailed geoneutrino analysis will be presented in a future publication. 

The geoneutrino flux at SNO+ is predicted using a methodology based on that in Ref.~\cite{Wipperfurth:2019idn,Wipperfurth:2018}. 
Assuming a total radiogenic heat production of 20~TW, and an average survival probability for the \nuebars 
($\langle P_{ee}\rangle \approx 0.55$), our model predicts 36.3$\pm$8.7~TNU and 9.7$\pm$2.3~TNU from the $^{238}$U and $^{232}$Th chains, respectively. The Terrestrial Neutrino Unit (TNU) is defined as the number of IBDs per year in 10$^{32}$ protons.  Thus, around 37 IBDs from geoneutrinos are expected in the total dataset.  
The unoscillated geoneutrino flux is unconstrained in this analysis, while the U/Th ratio is constrained to $3.8\pm1.3$, based on our model. 
Average oscillations are applied so that the flux is scaled by $\langle P_{ee} \rangle = \sin^4{\theta_{13}}+\cos^4{\theta_{13}}(1-0.5\sin^2{2\theta_{12}}) $. 

\smallskip
\textit{\alphan Background}~~\textemdash~~
The dominant background is $\alpha + ^{13}$C $\to$ $^{16}$O$ + n$.  The rate depends on the cross section of the interaction, the natural abundance of $^{13}$C in the scintillator, and the rate of $\alpha$ decays, which are dominated by $^{210}$Po. The fiducial volume of 5.7~m reduces the contribution from $^{210}$Po implanted on the AV to negligible levels.  Within the fiducial volume, the rate was measured directly by fitting the energy peak of $^{210}$Po $\alpha$'s as a function of time. 

The \alphan process is characterized by a delayed neutron capture, while the prompt event can be from the de-excitation of $^{16}$O (at 6.05~MeV and 6.13~MeV), a 4.4-MeV $\gamma$ from the neutron scattering with $^{12}$C, or a continuum from multiple scatters of the neutron on protons, in addition to the residual energy deposited by the $\alpha$ before it interacted. The multiple scattering signal dominates and extends from below the prompt event threshold of 0.9~MeV up to around 3.5~MeV. 

The probability of a $^{210}$Po $\alpha$ undergoing an \alphan interaction is calculated by integrating the energy-dependent interaction cross section up to the $^{210}$Po $\alpha$ energy, and then multiplying by the $^{13}$C number density, as in Ref.~\cite{snoplus_partial_antinu}. A branching ratio uncertainty of 30\% is assumed for the ground state channel, namely multiple-proton scatters (89\% of total) and the 4.4-MeV $\gamma$ (2\% of total).  A 100\% uncertainty is used for the $^{16}$O excited state channel (9\% of total).  

\smallskip
\textit{\alphap Background}~~\textemdash~~
$^{214}$Bi undergoes beta decay with an endpoint energy of 3.3~MeV and is followed by a $^{214}$Po $\alpha$ decay of energy 7.8~MeV and lifetime 164~$\mu s$, which is similar to the neutron capture time. 
Usually, the $\alpha$ scintillation signal is quenched down to $\approx$0.8~MeV; however, 
some of its energy can be transferred to protons by scattering. Since quenching is smaller for protons, a tail extends from the $^{214}$Po $\alpha$ peak to higher energies and can overlap with the neutron-capture energy peak. 

This \alphap process is a potential background for IBD measurements in scintillator detectors and was recently identified by SNO+~\cite{sno+ppo} in short-lived high-background periods where $^{214}$Bi$^{214}$Po coincidences were supported by the ingress of $^{222}$Rn. Its rate scales with that of the usual $^{214}$Bi$^{214}$Po coincidence and was observed to decay away with a 3.8-day half-life, as expected for $^{222}$Rn. 

As in Ref.~\cite{sno+ppo}, data from high-background periods were used to model the $^{214}$Po spectrum based on the similar tail observed for $^{215}$Po $\alpha$'s from the $^{235}$U chain. The resulting delayed energy spectrum was associated with a prompt spectrum from the $^{214}$Bi beta decay.  The precise shape of the tail depends on the quenching of $\alpha$'s and protons, as well as cross sections, and further studies are underway to model it. 
A total of 7.5 $\pm$ 6.2 $^{214}$Bi$^{214}$Po \alphap coincidences are expected across both datasets, where a common systematic uncertainty of 83\% reflects variations induced by the tail construction procedure.

\smallskip
\textit{Atmospheric Neutrino Background}~~\textemdash~~
Interactions of atmospheric neutrinos with the scintillator can also produce signals with accompanying neutrons. Charged-current interactions often produce high-energy depositions that are removed by the 1-s veto and by requiring only one prompt and one delayed event. Neutral current quasi-elastic interactions can produce neutrons with energies $\mathcal{O}$(10)~MeV. 

The expected atmospheric neutrino rate was estimated using GENIE Neutrino Generator version 3.4.2~\cite{atm-genie}. The flux from Bartol~\cite{atm-bartol} in the Solar-max phase and at the SNO+ location served as input for neutrinos above 100~MeV, while the smaller component below 100~MeV was taken from Ref.~\cite{atm-fluka}. Neutrino oscillations were applied with the NuCraft package~\cite{atm-nucraft}. 

Applying all event selection criteria to the simulated sample resulted in an expectation of 4.4 coincidences in the full dataset. The systematic uncertainties of the different components in the calculation combine to a total uncertainty of 68\%. The predicted prompt energy spectrum was consistent with a uniform distribution between 0.9~MeV and 8.0~MeV, which is assumed in the analysis.

\begin{table}[tbp]\centering
\caption{Expected and fitted signal and background counts.  Expectations use oscillation parameters from Ref.~\cite{PDG2025} and show only systematic uncertainties (the geo-$\nu$ IBD rate is unconstrained). The first Fit column is SNO+ only, the second Fit column is with externally-constrained oscillation parameters, and the last Fit column is with the cut on the \alphan classifier. For the latter, the expected count sum is 198.  See text for details.}\label{tab:sum_sig_bkg}
\begin{tabular}{lcccc}
\toprule
\toprule
   & Expected & Fit & Fit (con.) & Fit \alphan cut \\
\toprule
Reactor IBD & $140\pm6$ & $120\pm5$ & $137\pm6$ & $130\pm6$ \\
Geo $^{238}$U IBD & 29 & $32\pm16$ & $31_{-16}^{+15}$  & $27\pm8$ \\
Geo $^{232}$Th IBD & 8 & $9\pm5$  & $9\pm5$ & $8\pm4$ \\
\alphan $p$-scatters & $80 \pm 24$ & $67\pm20$ & $63_{-19}^{+20}$ & $23_{-6}^{+7}$ \\
\alphan other & $12 \pm 10$ & $7\pm5$ & $7\pm5$ & $8_{-5}^{+4}$ \\ 
\alphap  & $7\pm 6$ & $6\pm6$ & $4_{-4}^{+6}$ & $2_{-2}^{+6}$ \\
Atmospheric $\nu$ & $4\pm 3$ & $4\pm3$ & $6_{-2}^{+3}$ & $4\pm2$ \\
\hline
Sum & 281 & 247 & 257 & 201 \\
\hline
Observed  & 246 & 246 & 246 & 185 \\
\bottomrule
\bottomrule
\end{tabular}
\end{table}

\smallskip
\textit{Spectral Analysis and Results}~~\textemdash~~
A total of 246 coincidences were selected in the two datasets. 
This is compared with the expected numbers of signals and backgrounds in Table~\ref{tab:sum_sig_bkg}.
The expectations in each dataset are 101 and 180 coincidences, which are consistent with the observations of 103 and 143, respectively.  
Additional information is provided for each dataset in Supplemental Material~\cite{supp}.  
The prompt energy distribution of the combined dataset is shown in Fig.~\ref{fig:Eprompt}.

Two independent fitter frameworks used extended $\log$-likelihood functions to fit the unbinned prompt energy spectrum.  Both frameworks used the same selected event samples and input predictions, and produced consistent results, one set of which is presented and summarized in Tables~\ref{tab:sum_sig_bkg} and \ref{tab:fit_results}.  
In addition to the two oscillation parameters, we fit the geoneutrino rate and U/Th ratio, and scalings to the cross section and branching ratios for the \alphan channels. The normalizations of these and the other backgrounds were treated as fully correlated across the two datasets with the same relative constraints, together with a common scale for the reactor predictions. 
The detector-related systematics, including the energy resolution, a linear energy scale, and a nonlinear scaling related to Birks' constant $k_B$, were fit and constrained with distinct uncertainties for each dataset (see Table~\ref{tab:datasets}).  
A linear energy scale systematic for protons was fit without constraint, due to the lack of a calibration.  

\begin{figure}[tbp]
    \centering
    \caption{Likelihood contours of $\Delta m^2_{21}$ vs. $\sin^2\theta_{12}$. The 1-D $\log$ likelihood curve is also shown separately for each variable, profiling over the other. Contours and 1-D curves for results from KamLAND~\cite{kamland_on_off}, JUNO \cite{JUNO:2025gmd}, and combined solar measurements~\cite{Super-Kamiokande_solar} are also shown.}
    \includegraphics[width=\columnwidth]{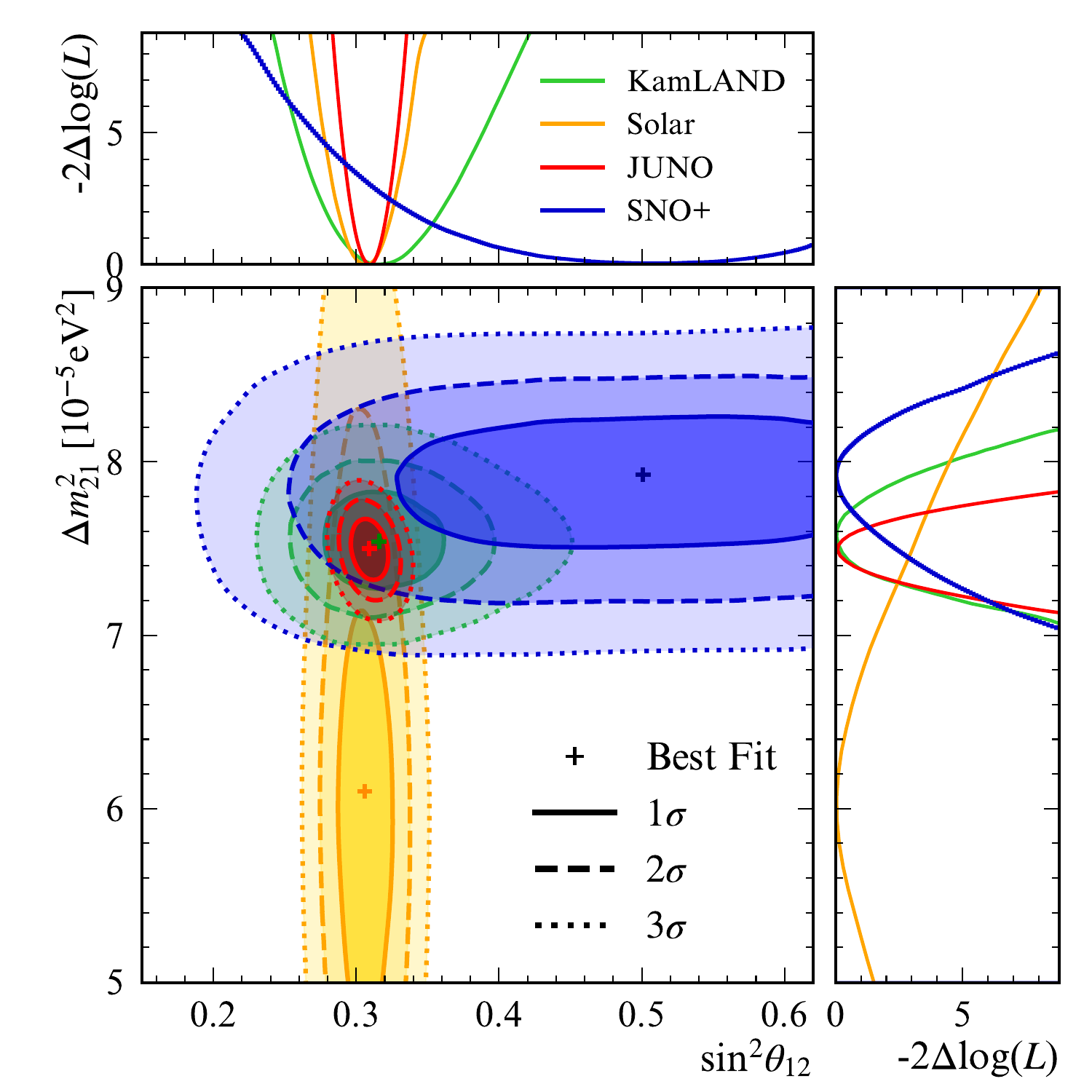}
    \label{fig:fit_results}
\end{figure}

The best-fit oscillation parameters are $\Delta m^2_{21} = (7.91^{+0.22}_{-0.25}) \times 10^{-5} \text{eV}^2$ and $\sin^2\theta_{12} = 0.505\pm0.133$. 
All best-fit detector systematic parameters are consistent with expectations and the resulting energy spectra are shown with the data in Fig.~\ref{fig:Eprompt}.  
The $\log$ likelihood distribution is shown in Fig.~\ref{fig:fit_results} as a function of the two oscillation parameters, fitting for all the other parameters at each point. 
The contribution of statistical uncertainty to the total uncertainty is estimated as the fit error when fixing all but the two oscillation parameters to their best-fit values, giving fractions of 96\% and 82\% for $\Delta m^2_{21}$ and $\sin^2\theta_{12}$, respectively. These estimates do not account for parameter correlations, but illustrate the predominance of the statistical uncertainty.

The SNO+ $\Delta m^2_{21}$ result is compatible with the KamLAND and JUNO results of $(7.54^{+0.19}_{-0.18}) \times 10^{-5}$ eV$^2$~\cite{kamland_on_off} and $(7.50\pm0.12) \times 10^{-5}$ eV$^2$~\cite{JUNO:2025gmd} to 1.2$\sigma$ and 1.5$\sigma$, respectively.  
A larger central value for SNO+ was hinted at by the previous SNO+ result of $(7.96^{+0.48}_{-0.42}) \times 10^{-5}$ eV$^2$~\cite{sno+ppo}, and is confirmed when datasets I and II are fit individually. 

The previous oscillation measurements from reactors are less sensitive to the mixing angle $\sin^2{\theta_{12}}$ than those from solar data. 
When the fit is repeated with the mixing angle constrained to the global result of $\sin^2{\theta_{12}} = 0.307\pm0.013$~\cite{PDG2025}, it returns a consistent value of $\Delta m^2_{21} = (7.87^{+0.26}_{-0.32})\times 10^{-5}$ eV$^2$. 
The combined solar result of $\Delta m^2_{21} = (6.10^{+0.95}_{-0.81}) \times 10^{-5}$~eV$^2$~\cite{Super-Kamiokande_solar} is consistent with the result from SNO+ at 1.9$\sigma$. 

The impact of SNO+ data on the global measurement of the oscillation parameters is probed by repeating the fit with both oscillation parameters constrained to the PDG 2025 global values~\cite{PDG2025} (which preclude JUNO's results), using Gaussian constraints.
This results in a very slight increase in the mixing angle $\sin^2{\theta_{12}} = 0.310\pm0.012$, and a moderate increase in the mass-squared difference, from $(7.50\pm0.19) \times 10^{-5}$ eV$^2$ to $(7.59\pm0.17) \times 10^{-5}$ eV$^2$.  
The global landscape of measurements of $\Delta m^2_{21}$ and $\sin^2{\theta_{12}}$ is shown in Fig.~\ref{fig:fit_results}.  

When constraining the oscillation parameters, the geoneutrino IBD rate fits to $50\pm25$ TNU with a U/Th ratio of $3.54_{-1.37}^{+1.36}$. 
The uncertainty in the geoneutrino rate is dominated by statistical uncertainty and a correlation of -0.5 with the multiple-proton scattering component of the \alphan background.

\begin{table}[tbp]
    \centering
    \caption{\label{tab:fit_results}Best-fit values for oscillation parameters and geo-$\overline{\nu}$ IBD rate. 
    Results are reported with no external constraints on $\Delta m^2_{21}$ and $\sin^2\theta_{12}$, then with constraints, and finally with constraints and the cut on the classifier for \alphan proton-scattering events.  See text for details.}
\begin{tabular}{lccc}
\toprule
\toprule
            & Fit & Fit (con.) & Fit \alphan cut\\
\midrule
$\Delta m^2_{21}$ [$10^{-5} \text{eV}^2$]  & $7.91^{+0.22}_{-0.25}$  & $7.59\pm0.17$ & $7.55^{+0.17}_{-0.16}$ \\
$\sin^2\theta_{12}$  & $0.505\pm0.133$  & $0.310\pm0.012$ & $0.311\pm0.012$ \\
Geo-$\overline{\nu}$ [TNU]  & $51_{-25}^{+24}$  & $50\pm25$ & $48^{+14}_{-12}$ \\
Geo-$\overline{\nu}$ U/Th & $3.55_{-1.37}^{+1.36}$ & $3.54_{-1.37}^{+1.36}$ &  $3.23_{-1.49}^{+1.43}$ \\
\bottomrule
\bottomrule
\end{tabular}
\end{table}

We also present a third fit that uses the \alphan classifier to reduce the dominant \alphan background below 3.5~MeV.  
The lower plot of Fig.~\ref{fig:Eprompt} shows the 185 events (77 and 108 for datasets I and II) that remain after applying the \alphan classifier cut.  The data are fitted with four additional parameters to account for the relatively large systematic uncertainties from a first evaluation of the classifier using the AmBe source: two global scalings for the IBD selection efficiencies in each dataset, and two energy-dependent scalings for the multiple-proton scattering efficiencies. 

\begin{figure}[htbp]
\caption{\label{fig:classifier} Distributions of the \alphan classifier value for prompt events with energy below 3.5~MeV, for the full dataset.
Classifier values are shifted so that the cut threshold for each dataset is at 0, denoted with a vertical dashed line. 
}
\includegraphics[width=1.0\columnwidth]{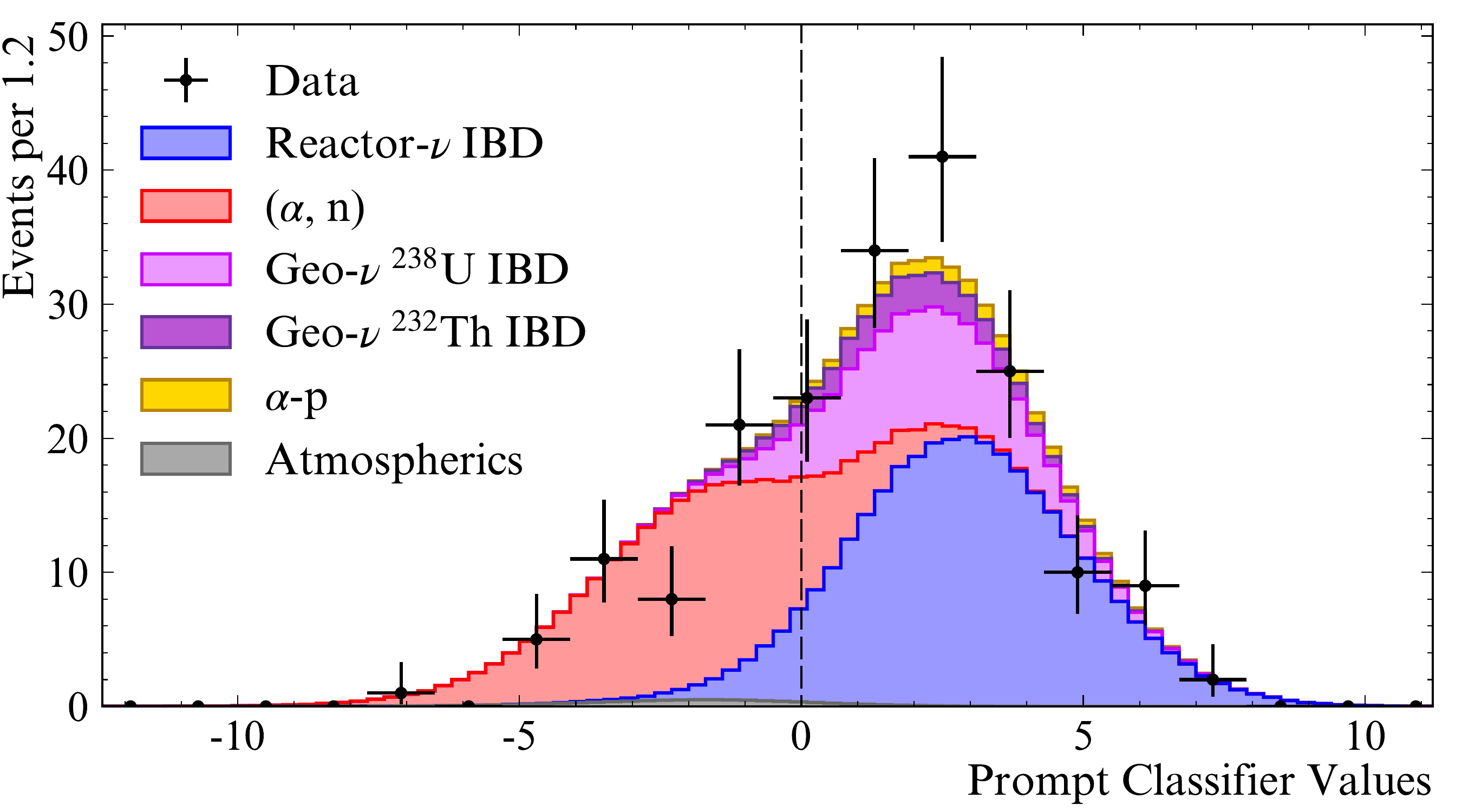}
\end{figure}

The fit with the \alphan classifier cut produces a geoneutrino measurement with a similar central value and a reduced uncertainty: $48^{+14}_{-12}$ TNU. 
The correlation of -0.5 with the \alphan background seen in the fit without the classifier is now more evenly distributed between the \alphap background and the remaining \alphan background. 
The normalizations of these backgrounds both exhibit residual correlations with the parameters introduced to describe the systematic uncertainties of the classifier. This suggests that further improvement to the geoneutrino measurement can be expected as the classifier-related systematic uncertainty is reduced.  The current performance of the \alphan classifier is summarized in Fig.~\ref{fig:classifier}.  
The oscillation parameters fitted when using the \alphan classifier are not taken as the final results of this study because the classifier currently introduces a large energy-dependent systematic uncertainty, which in principle could skew the fitted value of $\Delta m^2_{21}$.  In contrast, the geoneutrino flux is less sensitive to energy-dependent systematics.  The three different fits are summarized in Table~\ref{tab:fit_results}.

\smallskip
\textit{Summary and Outlook}~~\textemdash~~
SNO+ measures antineutrinos produced in the Earth and in nuclear reactors located hundreds of kilometers away. Its present measurement of $\Delta m^2_{21}=(7.91^{+0.22}_{-0.25})\times 10^{-5}$~eV$^2$ is dominated by statistical uncertainty and is approaching the measurement by KamLAND, which also used long-baseline reactor antineutrinos.  In addition, SNO+ has now detected geoneutrinos at a significance of 3.6$\sigma$. 

Interpreting geoneutrino measurements from KamLAND~\cite{KamLAND:2022vbm}, Borexino~\cite{Borexino:2019gps}, and now SNO+ and JUNO~\cite{JUNO:2025gmd}, includes evaluating the dominant component from the crust local to each experiment. The combination of results from these different sites will give a global picture of the Earth’s radiogenic heat production and can be used to extract the component from the mantle. At SNO+, the measured rate is 48~TNU with an uncertainty around 27\%; with this improved precision, a dedicated global geological interpretation is underway. 

Previous measurements of $\Delta m^2_{21}$ using solar neutrinos and long-baseline reactor antineutrinos are compatible with each other to about 1.5$\sigma$. 
The current result from SNO+ slightly increases this difference to 1.6$\sigma$.  
Fitting SNO+ data with the previous results as constraints provides new global best-fits of $\Delta m^2_{21} = (7.59\pm0.17) \times 10^{-5} \text{eV}^2$ and $\sin^2\theta_{12} = 0.310\pm0.012$. 

The measurement precision of several oscillation parameters, including $\Delta m^2_{21}$ and $\sin^2\theta_{12}$, is now being improved by the JUNO experiment~\cite{JUNO:2025gmd}, which is acquiring data with a larger detector (25~kt) and shorter baselines to reactors (53~km). The SNO+ Collaboration will continue to improve its result, which is complementary as the antineutrinos oscillate over greater distances and the analysis is subject to a different dominant background. 
At the same time, SNO+ will continue to collect data and characterize the models of \alphan and \alphap interactions, which are important in low-background particle physics experiments.

\begin{acknowledgments}
Capital funds for SNO\raisebox{0.5ex}{\tiny\textbf{+}} were provided by the Canada Foundation for Innovation and matching partners: 
Ontario Ministry of Research, Innovation and Science, 
Alberta Science and Research Investments Program, 
Queen’s University at Kingston, and 
the Federal Economic Development Agency for Northern Ontario. 
This research was supported by 
{\it Canada: }
the Natural Sciences and Engineering Research Council of Canada, 
the Canadian Institute for Advanced Research, 
the Arthur B. McDonald Canadian Astroparticle Physics Research Institute; 
{\it U.S.: }
the Department of Energy (DOE) Office of Nuclear Physics, 
the National Science Foundation and the DOE National Nuclear Security
Administration through the Nuclear Science and Security Consortium; 
{\it UK: }
the Science and Technology Facilities Council (STFC) and the Royal Society; 
{\it Portugal: } 
Funda\c{c}\~{a}o para a Ci\^{e}ncia e a Tecnologia (FCT-Portugal); 
{\it Germany: }
the Deutsche Forschungsgemeinschaft; 
{\it Mexico: }
DGAPA-UNAM and Consejo Nacional de Ciencia y Tecnolog\'{i}a; 
{\it China: }
the Discipline Construction Fund of Shandong University.  
We also thank SNOLAB and SNO\raisebox{0.5ex}{\tiny\textbf{+}} technical
staff; the Digital Research Alliance of Canada; the
GridPP Collaboration and support from Rutherford
Appleton Laboratory; and the Savio computational cluster
at the University of California, Berkeley. Additional long-term
storage was provided by the Fermilab Scientific Computing
Division.

For the purposes of open access, the authors have applied a Creative Commons Attribution licence to any Author Accepted Manuscript version arising. Representations of the data relevant to the conclusions drawn here are provided within this paper and its supplemental material.
\end{acknowledgments}

\bibliographystyle{apsrev}
\bibliography{References}

\end{document}